\documentstyle[12pt,psfig]{article}
\begin{document}
\centerline{\Large A Numerical Study on the Evolution of Portfolio Rules:}
\centerline{}
\centerline{\Large Is $CAPM$ Fit for Nasdaq?}
\centerline{Guido Caldarelli$^{1}$, Marina Piccioni$^1$ and 
Emanuela Sciubba$^{2}$}
\centerline{\em $^1$ INFM - Unit\`a di Roma 1 ``La Sapienza", P.le A. Moro 2,
00185 - Roma, Italy}
\centerline{\em $^2$ Tinbergen Institute Rotterdam, 
Burg. Oudlaan 50 NL - 3062 PA Rotterdam, The Netherlands}

\begin{abstract}
In this paper we test computationally the performance of $CAPM$ in an
evolutionary setting. In particular we study the stability of wealth
distribution in a financial market where some traders invest as prescribed
by $CAPM$ and others behave according to different portfolio rules. Our study
is motivated by recent analytical results that show that, whenever a
logarithmic utility maximiser enters the market, traders who either
``believe'' in $CAPM$ and use it as a rule of thumb for their portfolio
decisions, or are endowed with genuine mean-variance preferences, vanish in
the long run. Our analysis provides further insights and extends these
results. We simulate a sequence of trades in a financial market and: first,
we address the issue of how long is the long run in different parametric
settings; second, we study the effect of heterogeneous savings behaviour on
asymptotic wealth shares.
We find that $CAPM$ is particularly ``unfit'' for highly risky environments.
\end{abstract}

\smallskip


\section{Introduction}
A major part of research in financial economics is aimed at improving our
understanding of how investors make their portfolio decisions and hence of how 
asset prices are determined.
In this paper we contribute to that research adopting a dynamic approach, where 
investors' wealth endowments and portfolio rules evolve in time. 

We adopt here an evolutionary framework to address the issues of 
\textit{fitness} and \textit{survival} in financial markets, where traders differ in 
portfolio rules and/or savings behaviour. We build a simple framework that
allows us to simulate a long sequence of trades in a competitive financial 
market and to test computationally the asymptotic wealth distribution of traders 
who follow heterogeneous portfolio rules. In particular we concentrate our analysis 
on the interaction between two types of traders: traders who use
(the traditional version of) $CAPM$ as a rule of thumb for their portfolio decisions and 
traders who make portfolio choices maximising a logarithmic utility function. 

The choice of this particular focus of analysis is motivated by an 
old debate and a recent strand of literature. 
Despite the fact that its restrictive assumptions have often been criticised and 
its predictive power has been challenged by numerous contribution in empirical
finance (see for example Fama and French\cite{ff1,ff2}), 
the mean variance approach is a 
standard in financial economics, and its main corollary in asset pricing, 
the Sharpe-Lintner-Mossin Capital Asset
Pricing Model \footnote{See Sharpe \cite{sharpe}, Lintner \cite{lintner} and 
Mossin \cite{mossin}.}, has
been viewed as one of 
the ``major contributions of academic research in
the postwar era'' (Jagannathan and Wang \cite{jagannathan}, p.4). 
Since a seminal article by Kelly \cite{kelly},
financial economists and applied probability theorists have been debating on the 
normative appeal of logarithmic utility maximisation as opposed to the mean-variance 
approach in financial markets. 
Several contributors\footnote{For example: Latane \cite{latane}, Breiman 
\cite{breiman}, Hakansson \cite {hakansson}, Finkelstein and Whitley \cite{finkelstein}, 
Algoet and Cover \cite{algoet}.} have expressed their
dissatisfaction with the mean-variance approach and argued that a rational investor with 
a long time horizon should instead maximise the expected rate of growth of his wealth. 
The portfolio choice that this
type of behaviour implies is equivalent to that prescribed by a logarithmic utility 
function (the so called ``Kelly criterion''). Central to the debate was the claim that 
maximising a logarithmic utility function would be a ``more rational'' objective to 
follow for a trader with a long time horizon. This claim has been strongly
opposed by Merton and Samuelson \cite{merton} and Goldman \cite{goldman}, who stressed 
the obvious contradiction that lies in arguing that rational traders should maximise a 
given utility function, possibly different from their own. 

We believe that a useful 
contribution to this debate comes by the adoption of an
evolutionary technique. By showing that logarithmic utility maximisers accumulate 
more wealth than $CAPM$-traders we certainly do not prove that they are more rational 
(or happier) than their mean-variance
opponents. However we may gain a better understanding of the reasons 
for the support of logarithmic utility
maximisation that originated the debate. 

We do not attempt here a review on the literature on evolution and market behaviour.
Instead we concentrate on the more self-contained 
strand of literature to which this paper specifically contributes. In particular, 
we focus on the literature that aims at studying long run
financial market outcomes as the result of a process akin to natural selection. 
In a seminal paper Blume and Easley \cite{blume} develop an evolutionary model of 
a financial market and define notions of dominance,
survival and extinction based on the asymptotic behaviour of traders' wealth shares. 
They provide us with the most general analytical result in this strand of literature. 
They show that, if all traders save at the same
rate and under some uniform boundedness conditions on portfolios, then there exists 
one globally fittest portfolio rule which is prescribed by logarithmic utility 
maximisation. Namely, if there is a logarithmic
utility maximiser in the economy, he will dominate, determine asset prices 
asymptotically and drive to extinction any other trader that does not behave, at least 
in the long run, as a logarithmic utility maximiser.

In a different setting, where investors choose their investment rates endogenously 
(choosing an optimal consumption stream at the same time as an optimal sequence of 
portfolios), Sandroni \cite{sandroni} provides analytical results that seem to weaken 
Blume and Easley's findings. He shows that, provided that
agents' utilities satisfy Inada conditions (and that agents' beliefs are correct), 
then all traders survive.  Clearly Inada conditions are crucial here as they require 
marginal utilities to go to infinity whenever
the consumption level approaches zero. A rational trader that can also choose 
his investment intensity would avoid extinction by suitably modifying his consumption 
pattern. 

Sciubba \cite{sciubba} compares the
relative fitness of logarithmic and mean-variance preferences. Mean-variance 
preferences do not satisfy Inada conditions and prescribe portfolio weights which do 
not necessarily display uniform boundedness
properties, so that none of the two previous frameworks directly applies. She shows that 
when logarithmic utility maximisers invest at the same rate as mean-variance investors, 
logarithmic traders dominate, determine asset prices asymptotically and drive to extinction 
those agents who are either endowed with
mean-variance preferences, ore use their theoretical predictions ($CAPM$) as a rule of thumb. 
One of the sections in Sciubba \cite{sciubba} is indeed devoted to the study of the 
role of heterogeneous savings rates in the dynamics of wealth accumulation. 
However, the stochastic process becomes so complex in this case, that the
few results that are obtainable analytically are very weak. This is where we believe that 
numerical computation can provide us with more insights into the problem. 

Our results show that, even when we allow for heterogeneity in savings rates, dominance of 
logarithmic utility maximisers proves robust, at least to a certain extent. In particular, 
we show that logarithmic traders can ``afford" to save at a lower rate than $CAPM$ traders and 
still dominate financial markets. When logarithmic traders dominate, the wealth share of $CAPM$ 
traders converges to zero exponentially fast. In particular, we find that the more 
risk-averse $CAPM$ traders are, the faster they vanish.
Clearly when $CAPM$ traders display a savings intensity that is much higher than their 
opponents, then - as one would expect - they indeed dominate and drive to extinction 
logarithmic utility maximisers. We identify the threshold in the savings rates differential 
such that this ``fate reversal" occurs and find that it is increasing in the variance of the 
dividend stream.
If the dividend stream is very volatile, then the advantage of logarithmic utility maximisers 
with respect to $CAPM$ traders in terms of portfolio selection is higher. Hence $LOG$ traders can 
``afford" to save less and still dominate the market. Symmetrically, the fitness of $CAPM$ 
proves particularly low in environments characterised by high volatility. Should we trust 
what $CAPM$ prescribes when investing on Nasdaq? Probably not.

\section{The Model} 
A detailed description of the model is to be found in Sciubba \cite{sciubba}. 
Here we summarise its main features.
Time is discrete and indexed by $t=0,1,2,...$. 
There are $S$ states of the world indexed by $s=1,2,...,S$, one of which will occur at each date. 
States follow an i.i.d. process with
distribution $p=(p_{1},p_{2},...,p_{S})$ where $p_{s}>0$ for all $s$. 
There is a finite set of assets, with the same dimensionality
of the set of states of nature. With little abuse of notation we index
both assets and states with the same letter.
Asset $s\in \left\{ 1,2,...,S\right\} $ pays a strictly positive dividend $d_{s}>0$ when state 
$s\in \left\{ 1,2,...,S\right\} $ occurs and $0$
otherwise. At each date there is only one unit of each asset available, so that $d_{s}$ 
will be the total wealth in the economy at date $t$ if state $s$ occurs. 
This wealth will be distributed among traders
proportionately according to the share of asset $s$ that each of them owns. 
Let $\rho _{st}$ be the market price
of (one unit of) asset $s$ at date $t$. There is a heterogeneous population of 
long-lived agents in this
economy, indexed by $i\in \left\{ 1,2,...,I\right\} $. Each agent is 
characterised by an investment rule and an
initial wealth endowment. Agent $i$'s investment rule is 
$\left\{ \delta_{t}^{i},\alpha _{t}^{i}\right\}_{t=0}^{\infty }$ 
where $\delta _{t}^{i}$ denotes agent $i$'s 
investment rate at date $t$ (i.e. the
percentage of wealth endowment at date $t$, $e_{t}^{i}$, that $i$ 
invests at date $t$) and $\alpha_{t}^{i}$ is a vector that describes agent 
$i$'s portfolio choice at date $t$ (i.e. the vector of portfolio
weights in the $S$ available assets $\left\{ \alpha _{st}^{i}\right\}_{s=1}^{S}$ 
for agent $i$ at date $t$).
Agent $i$'s initial wealth endowment is denoted by $e_{0}^{i}$. The tuple 
$\left( e_{0}^{i},\left\{ \delta_{t}^{i},\alpha_{t}^{i}\right\}_{t=0}^{\infty }\right)$ 
constitutes a complete description of agent $i$. At each date $t$, agent $i$ invests 
a portion $\delta _{t}^{i}$ of his wealth endowment $e_{t}^{i}$, in the $S$
available assets\footnote{We are in fact assuming that agent $i$ ``eats'' the rest 
of his endowment so that
whatever is not invested, $\left( 1-\delta_{t}^{i}\right) e_{t}^{i}$, is 
consumed by agent $i$.}. We denote by $w_{t}^{i}$ the total amount invested by trader 
$i$ at date $t:$ 
\begin{equation} 
w_{t}^{i}=\delta_{t}^{i}e_{t}^{i} 
\end{equation} 
Aggregate total investment will be equal to aggregate expenditure in
assets. Since there is only one unit of each asset available, then clearly: 
\begin{equation}
\sum_{s=1}^{S}\rho_{st}=\sum_{i=1}^{I}w_{t}^{i}=w_{t} 
\label{rel} 
\end{equation} 
Equation (\ref{rel})
provides us with a convenient normalisation for prices. 
We can, in fact, call $\pi _{st}$ the normalised
market price of asset $s$ at date $t$ and define it as follows: 
\begin{equation} 
\pi _{st}\equiv \frac{\rho_{st}}{\sum_{s=1}^{S}\rho _{st}}=\frac{\rho _{st} }{w_{t}} 
\end{equation} 
Finally define: 
$\pi_{t}\equiv \left( \pi _{1t},\pi _{2t},...,\pi _{St}\right)$. 
In equilibrium, prices must be such that markets
clear, i.e. total demand equals total supply. Agent $i$'s demand for asset $s$ will 
be equal to agent $i$'s expenditure in asset $s$, $\alpha_{st}^{i}\,w_{t}^{i}$, 
divided by the market price of asset $s$, $\rho
_{st}$. In the aggregate: 
\begin{equation} 
\sum_{i=1}^{I}\frac{\alpha _{st}^{i}\,w_{t}^{i}}{\rho _{st}}=1
\label{equilibrium} 
\end{equation} 
Rewriting equation (\ref{equilibrium}) we get: 
\begin{equation} 
\rho_{st}=\sum_{i=1}^{I}\alpha _{st}^{i}\,w_{t}^{i} 
\label{price equation} 
\end{equation} 
and using our
price normalisation: 
\begin{equation} \pi _{st}=\sum_{i=1}^{I}\alpha _{st}^{i}\,\varepsilon _{t}^{i}
\label{normp} 
\end{equation} where $\varepsilon _{t}^{i}$ denotes the investment share of agent $i$ at
date $t$: 
\begin{equation} 
\varepsilon _{t}^{i}=\frac{w_{t}^{i}}{w_{t}} 
\end{equation} 
and also measures the ``presence'' of trader $i$ in financial markets and his ``importance'' in 
determining asset prices. 

\subsection{The Dynamics\label{subsec}} 
We now ask what is the period to period dynamics of
trader $i$'s investment share. This will allow us to follow the evolution of his 
``presence'' in financial markets and of his ``importance'' in determining market 
outcomes, namely asset prices. Suppose that state
$s$ occurs at date $t$: asset $s$ pays a dividend $d_{s}$ 
while all other assets pay zero. The total wealth 
in the economy, $d_{s}$, gets distributed to traders according to the share of 
asset $s$ that each of them owns. In particular, the investment income of trader $i$ 
- that also constitutes his wealth endowment for period $t+1$ - is equal to the following: 
\begin{equation} 
e_{t+1}^{i}=\frac{\alpha _{st}^{i}w_{t}^{i}}{\rho_{st}}d_{s} 
\end{equation} Aggregate wealth endowment at date $t+1$ is equal to the total payout of
asset $s$: 
\begin{equation} 
e_{t+1}=\sum_{i=1}^{I}e_{t+1}^{i}=\sum_{i=1}^{I}
\frac{\alpha_{s}^{i}w_{t}^{i}}{\rho _{st}}d_{s}=d_{s} 
\label{ee} 
\end{equation} In period $t+1$ trader $i$ invests a
fraction $\delta_{t+1}^{i}$ of his wealth endowment: 
\begin{equation} w_{t+1}^{i}=
\delta_{t+1}^{i}e_{t+1}^{i}=\delta _{t+1}^{i}\frac{\alpha _{st}^{i}d_{s}}{\rho _{st}}w_{t}^{i} 
\label{w}
\end{equation} 
In the aggregate: 
\begin{equation}
w_{t+1}=\sum_{i=1}^{I}w_{t+1}^{i}=\sum_{i=1}^{I}\delta_{t+1}^{i}
\frac{\alpha_{st}^{i}w_{t}^{i}}{\rho_{st}}d_{s}=\sum_{i=1}^{I}
\delta_{t+1}^{i} \frac{\alpha _{st}^{i}w_{t}^{i}}{\rho_{st}}e_{t+1} 
\label{ww} 
\end{equation} 
In order to formulate (\ref{w}) in terms of wealth shares, it is useful to define the 
market average investment rate $\delta _{t+1}$ as:
\begin{equation} 
\delta_{t+1}=\sum_{i=1}^{I}\delta_{t+1}^{i}\frac{\alpha_{st}^{i}w_{t}^{i} }
{\rho_{st}} 
\end{equation} 
so that (\ref{ww}) can be rewritten as: 
\begin{equation} 
w_{t+1}=\delta_{t+1}e_{t+1} 
\end{equation} 
Dividing (\ref{w}) by (\ref{ww}) we obtain investment shares:
\begin{equation} 
\varepsilon _{t+1}^{i}=\frac{w_{t+1}^{i}}{w_{t+1}}=\frac{\delta_{t+1}^{i}\alpha
_{st}^{i}d_{s}}{\delta_{t+1}\rho _{st}e_{t+1}}w_{t}^{i} 
\end{equation} 
Finally, using (\ref{ee}) and
our price normalisation, we obtain: 
\begin{equation} 
\varepsilon _{t+1}^{i}=\frac{\delta _{t+1}^{i}\alpha_{st}^{i}}
{\delta _{t+1}\pi _{st}}\varepsilon _{t}^{i} \label{e} 
\end{equation} 
Equation (\ref{e}) describes
the period to period dynamics of the investment share of trader $i$. 
It implies that the ``impact'' of trader
$i$ on financial markets follows a fitness-monotonic dynamic: trader $i$'s investment share 
will increase if he invests more than the market on average and if he scores, 
with his investments, a payoff which is higher than the average population payoff. 
In fact, if state $s$ occurs at date $t$, $\alpha _{st}^{i}$ and $\pi _{st}$ give 
us a measure of trader $i$'s payoff (his bet on the lucky asset) 
and of the average population payoff respectively. 
Following the definition given by Blume and Easley 
\cite{blume}, in order to establish whether trader $i$ survives or vanishes,  
we consider the asymptotic value of his investment share and we check whether 
it is bounded away from zero or not. When trader 
$i$'s investment share converges \footnote{Here and in what follows 
whenever we refer to convergence, we mean almost sure convergence.} 
to zero, his significance in determining 
market outcomes vanishes and he effectively disappears from the market. 
When trader $i$'s investment share stays bounded away from zero, then he contributes 
to determine asset prices (also asymptotically) and we will say that, as a trader, 
he survives. The process described by eq.(\ref{e}) is too complex to be 
studied analytically in a very general case\footnote{ I.e. without imposing some of 
the restrictions required, for example, by \cite{blume}, \cite{sandroni} or 
\cite{sciubba}.}. In fact, even under the restriction that all shocks are i.i.d., 
asset prices and market savings rates depend on the whole sequence 
of past history, and the law of large numbers does not automatically help us 
with a solution. As we argued in the introduction, this is why we believe that numerical 
analysis can help to obtain further insights into the problem. 

\subsection{Types of Traders} 
As in Sciubba \cite{sciubba},
we will consider the interaction between two different types of traders. 
The first type of agents is given by 
investors who believe in $CAPM$ and use it as a rule of thumb
\footnote{We can think of them as of traders who have been educated in 
business schools. 
They have been taught the model so well in their
finance courses that they believe it really works.}: 
at the beginning of each time period, they observe 
payoffs and market prices and work out the composition of the market 
and the risk-free portfolios. Finally,
according to their degree of risk aversion they choose their preferred 
combination between the two. At date
$t$, $CAPM$ investors choose $\gamma _{t}\in \left[ 0,1\right]$ and 
invest in asset $s$ a portion $\alpha_{st}^{CAPM}$ of their portfolio such that: 
\begin{equation} 
\alpha_{st}^{CAPM}=\gamma_{t}\alpha_{st}^{F}+\left( 1-\gamma_{t}\right)\alpha_{st}^{M} 
\label{capm1} 
\end{equation} 
where:
\begin{equation} 
\alpha _{st}^{F}\equiv \frac{\rho _{st}/d_{s}}{\sum_{z}\rho_{zt}/d_{z}}= 
\frac{\pi_{st}/d_{s}}{\sum_{z}\pi _{zt}/d_{z}} \label{f} 
\end{equation} 
and 
\begin{equation} 
\alpha_{st}^{M}\equiv \frac{\rho_{st}}{\sum_{z}\rho_{zt}}=\pi_{st}. 
\label{m} 
\end{equation} 
The second
type of traders is given by investors who are endowed with a logarithmic utility 
function (type $LOG$) and who actually maximise the growth rate of their wealth share 
and invest according to a ``simple'' and time
invariant portfolio rule: 
\begin{equation} 
\alpha _{st}^{LOG}=p_{s} 
\label{logge}
\end{equation} 
More generally, at each
date $t$, a rational trader $i$ will choose $\left\{ \alpha _{st}^{i}\right\} _{s=1}^{S}$ so as to maximise:
\begin{equation} 
\sum_{s=1}^{S}p_{s}u^{i}\left( \frac{\alpha_{st}^{i}w_{t}^{i}}
{\rho_{st}} d_{s}\right) 
\label{prob} 
\end{equation} subject to the constraint that investment expenditure at each date
is less than or equal to the amount of wealth saved in the previous period. 
If $ u^{i}\left( \cdot \right) $ is
logarithmic, it follows that
\footnote{Note that we are assuming that traders know the probability
distribution $p$ over the state space $S$. In a more general framework, 
a trader who displays a 
logarithmic utility function bets his beliefs.} $\alpha _{st}^{L}=p_{s}$. 
We will denote by 
$\delta_{t}^{CAPM}$ and $\delta_{t}^{LOG}$ the investment 
rates of traders who believe in $CAPM$ and logarithmic utility maximisers 
respectively.

\section{Computer Simulation and Numerical Results} 
Recall eq.(\ref{normp}) where normalised prices are expressed as a weighted average
of portfolio rules. In the setting that we consider, where $CAPM$ and $LOG$ traders
interact, eq.(\ref{normp}) becomes
\begin{equation}
\pi_{st}=\varepsilon_t \alpha^{LOG}_{st}+(1-\varepsilon_t)\alpha^{CAPM}_{st}
\label{prezzi}
\end{equation}
where $\varepsilon_t$ denotes the investment share of $LOG$ traders and
$(1-\varepsilon_t)$ denotes the investment share of $CAPM$ traders.
By substituting eq.(\ref{capm1}) and eq.(\ref{logge}) in eq.(\ref{prezzi}) one has
\begin{equation}
\pi_{st}=\varepsilon_t p_s + 
(1-\varepsilon_t)(\gamma\frac {\pi_{st}}{d_s}
\frac 1 {\sum_{z=1,N}\frac{\pi_{zt}}{d_z} }
+ (1-\gamma)\pi_{st}).
\label{muuu}
\end{equation}

We solve the above equation through iteration by a numerical technique called 
``relaxation''. Namely, we start from a trial value for $\pi_{st}$, 
we compute a new value through the above 
equation and then  we iterate this procedure until a fixed point is 
reached.
In other words, if we call $\tau$ the time of the iteration our 
numerical code computes the quantity 
\begin{equation}
\pi_{st}^{\tau+1}=\varepsilon_t p_s + 
(1-\varepsilon_t)(\gamma\frac {\pi_{st}^{\tau}}{d_s}
\frac 1 {\sum_{z=1,N}\frac{\pi_{zt}^{\tau}}{d_z} }
+ (1-\gamma)\pi_{st}^{\tau})
\label{mu2}
\end{equation}
and then it looks for the fixed point solution $\pi_{st}^{\infty}$ that is
achieved for all the $\tau > \tau^*$ for which 
$|\pi_{st}^{\tau^*+1}-\pi_{st}^{\tau^*}|$ 
is less than a tolerance parameter fixed in advance for all
the assets $s$. This fixed point is the numerical solution of eq.(\ref{mu2}).
Since it has been demonstrated that this solution exists and is unique 
\cite{sciubba}, it is possible to show that the above method gives a numerical 
value of the solution with a desired precision bounded only by computation time.
In our simulations we stop the iteration when the relative precision of asset 
prices is lower than $10^{-5}$.
At this point we run two sets of simulations. 
The first set aims at detecting the time of convergence of 
the stochastic process given by the wealth shares.
The second set of simulations aims at checking the robustness of $LOG$ dominance results to 
heterogeneity in savings rates. 

The first numerical check is devoted to study the average time one 
can expect $CAPM$ traders to survive in a competition with the others.
We run a Fortran code that simulates a real evolution in a 
market with $CAPM$ traders and $LOG$ traders. 
We consider a market where 100 assets are repeatedly traded, so that $S=100$. 
We assume that the probability distribution over the states of nature 
is uniform, so that $p_s=1/S, \, \forall s$. We assume that dividends are randomly drawn 
from a normal distribution $N(\mu, \sigma)$ with the mean to variance ratio large enough to 
guarantee that virtually every asset pays a positive payoff. We set the initial investment
shares for the two types of traders to be equal. 
None of the qualitative results that we obtain in the simulation are driven by our choice 
of parameters and probability distributions.
We run our simulations under the assumption (as in Sciubba \cite{sciubba} and Blume and 
Easley \cite{blume}) that both types of traders save at the same rate.
Time step after time step the code generates 
pseudo-random numbers that identify states of nature. 
After each draw, prices are computed through ``relaxation'', 
the dividends are distributed and the investment shares of the traders are refreshed. 
Eventually the investment share of $CAPM$ traders becomes lower than a certain 
threshold, sufficiently small such that
we can conclude that $CAPM$ traders are extinct. 
We then store in the computer the time at which this threshold is reached and we start
again with a new realisation of the same market.
Given the same initial conditions in the market, we collect up to $15000$ different 
realisations of this competition between traders.
The various realisations differ because of the randomness in the draw of the states
of nature.
In all the runs $CAPM$ traders see their investment shares reduced 
until extinction takes place. This result confirms the theoretical findings. 
Furthermore,  through the extinction times recorded by the computer code, 
we are also able to measure the
density function describing the probability that $CAPM$ traders survive 
up to a time $t$ in a
competition with $LOG$ traders. The unit of time is given by the draw of a
state of nature. 
That is, $t=5$ means that after the initial condition $5$ states of nature have been 
drawn in the market, and the $5$ corresponding assets paied their dividends.
The main result is that the density function is exponential. 
The probability for a $CAPM$ trader to survive decreases with time according to
\begin{equation}
P(t)dt=Ae^{-Ct}dt.
\end{equation}
We also run different simulations by changing the parameter $\gamma$ of the model, that  
measures the risk-aversion of $CAPM$ traders.
The result is that the more risk-averse the $CAPM$ traders are, the faster their wealth 
share converges to zero. 
The economic intuition for this result lies in the fact that a higher degree of risk-aversion 
implies that the portfolio of $CAPM$ is tilted towards the risk-free rather than towards 
the market portfolio, where the most successful trading rules are represented. 
In this setting, a $CAPM$ trader with an extremely low risk-aversion, and hence a 
value of $\gamma$ approximately equal to zero, would indeed survive.

Our simulations show that exponential decay is robust with respect to the 
values of $\gamma$ used.
The functional form that can be hypothesised for such a decay is of the form
\begin{equation}
P(t)=Ae^{-Bt\gamma^2} 
\end{equation}
One can test this assumption by rescaling the different data.
Namely in each simulation, if we multiply the value of time by
the value of $\gamma^2$ used in that run, we should obtain different functions $P'(t')$
(where $t'=t\gamma^2$) all behaving in the same way.
In particular these data obtained through simulations with different values of $\gamma$ 
should collapse together. This technique note as ``collapse plot'' is shown in 
Fig.1. In the upper part of the figure we plot the data relative to
$\gamma=0.5,0.7,0.8,0.9,1$ testing the good validity of our assumption.
We also run another set of simulations, by changing the threshold at which
the $CAPM$ trader is considered extinct. By passing to a threshold of 
$0.5\%$ of the total wealth shares from the $5\%$ previously used, we 
noticed as expected a shift of the distribution to longer times.
Nevertheless, qualitatively we obtain the same behaviour shown in the lower part
of Fig.1. In both cases (above and below) the extimated exponential distribution
is indicated by a dashed line.

In the second set of simulations
we consider the situation in which the savings rates used by the two 
types of traders differ.
In particular, we ask if $LOG$ traders can survive and eventually dominate 
over $CAPM$ traders that save at a higher intensity.
We run simulations of the financial market described in the previous section, in this case 
under the assumption that traders have heterogeneous savings rates. 
Our aim is to check whether $LOG$ dominance results are robust 
when $CAPM$ traders save at a higher rate than logarithmic utility maximisers. 
As in the previous set of simulations, we consider a market where 100 assets are 
repeatedly traded, such that $S=100$. 
Again, we assume that the probability distribution over the states of nature is uniform, 
so that $p_s=1/S, \, \forall s$. 
We assume that dividends are randomly drawn from a normal distribution 
$N(\mu, \sigma)$ with the mean to variance ratio large enough to guarantee that 
virtually every asset pays a positive payoff. 
We set the initial wealth shares for the two types of traders to be equal. 
Finally we normalise the savings rate of $CAPM$ traders to be equal to one.

Fig. 2 shows the lowest value for logarithmic utility maximisers' savings rate that 
still allows $LOG$ traders to 
dominate (averaged across $1000$ different simulations), for different values of the variance 
$\sigma$ of the distribution of dividends, which measures the volatility of the dividend stream. 
More particularly,  we introduce the quantity $\Delta$ 
representing the maximum difference in saving rates between $CAPM$ and $LOG$ 
that still allows $LOG$ traders to dominate.
Also in this case we apply the technique of the collapse plot in order to show that
the threshold at which $LOG$ traders commence to dominate depends on the ratio $\sigma/\mu$.
We can interpret the graph in figure 2 as dividing the $\sigma$, $\Delta$ plane in two regions: 
in the area (a) below the graph, logarithmic utility maximisers dominate and drive to 
extinction $CAPM$ traders; in the area (b) above the graph, 
$CAPM$ traders dominate and drive to extinction logarithmic utility maximisers.
From our results it appears that the dominance of logarithmic utility maximisers is robust 
to heterogeneity in savings rates, at least to a certain extent. 
In fact we obtain that $LOG$ traders dominate and drive $CAPM$ 
traders to extinction even when the latter display a higher savings rate 
(provided that it is not much higher).
This difference provides us with an immediate measure of the advantage of $LOG$ 
traders with respect to $CAPM$ traders in terms of portfolio rules. We find that the 
higher the volatility of the dividend stream, 
the higher is the advantage of logarithmic utility maximisers over $CAPM$ traders. 
In particular, the threshold level for the $LOG$ traders' savings rate appears to be 
a function of the mean to variance ratio of the probability distribution of the 
dividends that assets pay.
The economic intuition of this result lies in the fact that,
with a large variance of dividends, the behaviour prescribed by a
logarithmic utility function differs greatly from the behaviour
prescribed by CAPM. Think of the following example. Suppose  that
investors can place bets on two lotteries that pay positive prizes with
equal probability: lottery "Rich" pays 1 million dollars with
probability 1/2 and lottery "Poor" pays 1 single dollar whenever lottery
"Rich" does not pay. A LOG investor would place equal bets on both
lotteries. On the contrary, a CAPM trader (and any trader, in fact, that
maximises the expected value of an increasing function of his wealth
level, rather than rate of growth) would place a higher bet on the
"Rich" lottery than on the "Poor" one. The divergence between his bets
would be increasing in the difference between the two lotteries' prizes.
As a result, the distance between the bets that LOG investors and CAPM
investors would place is also increasing in the discrepancy between
lotteries' payoffs. A higher variance of dividends implies that, in
different states of nature, very high payoffs are paid as well as very
low ones, and results in a higher divergence of portfolio choices
between the two types of investors.

\section{Conclusion}
In this paper we test computationally the performance of $CAPM$ in an evolutionary setting. 
In particular we study the asymptotic wealth distribution across two types of traders 
that compete on financial markets: traders who invest as prescribed by $CAPM$ and traders 
who maximise the expected value of a logarithmic utility function of wealth. 
Our study provides further insights and extends 
some recent analytical results (see \cite{sciubba}) that prove that the wealth share of 
$CAPM$ traders converges almost surely to zero, when a logarithmic utility maximiser 
with a savings rate at least as large as the savings rate of $CAPM$ traders, enters the 
market.
We run two sets of simulation addressing two related, but separate, issues. 
First, we look at the time of convergence of the stochastic process given by the wealth 
share dynamics. We find that, when savings rates are identical across the two types of 
traders, the wealth share of $CAPM$ investors decreases 
exponentially fast towards zero. We also find that the degree of risk-aversion of $CAPM$ 
traders has a role in determining the speed of convergence: the more risk-averse the 
$CAPM$ traders, the faster their wealth share converges to zero.
Second, we check the robustness of the analytical result in Sciubba \cite{sciubba} 
to heterogeneity in savings rates. We find that $LOG$ investors dominate even when their 
savings rate is lower (but not too much lower) than the savings rate of $CAPM$ traders. 
We compute the maximum difference between the savings 
rates of the two types of investors that still allows $LOG$ traders to dominate and drive 
to extinction all those who choose their portfolio according to what $CAPM$ prescribes. 
We argue that the difference between savings rates so computed can serve as a measure 
of the fitness of logarithmic utility maximisation with 
respect to $CAPM$ and we find that it is increasing in the variance of the dividend stream. 
Our results seem to suggest that, from an evolutionary perspective, if it is true that 
$CAPM$ could perform almost as satisfactorily as logarithmic utility maximisation in markets 
with low volatility, it proves particularly unfit for highly risky environments.

\begin{figure}
\centerline{\psfig{file=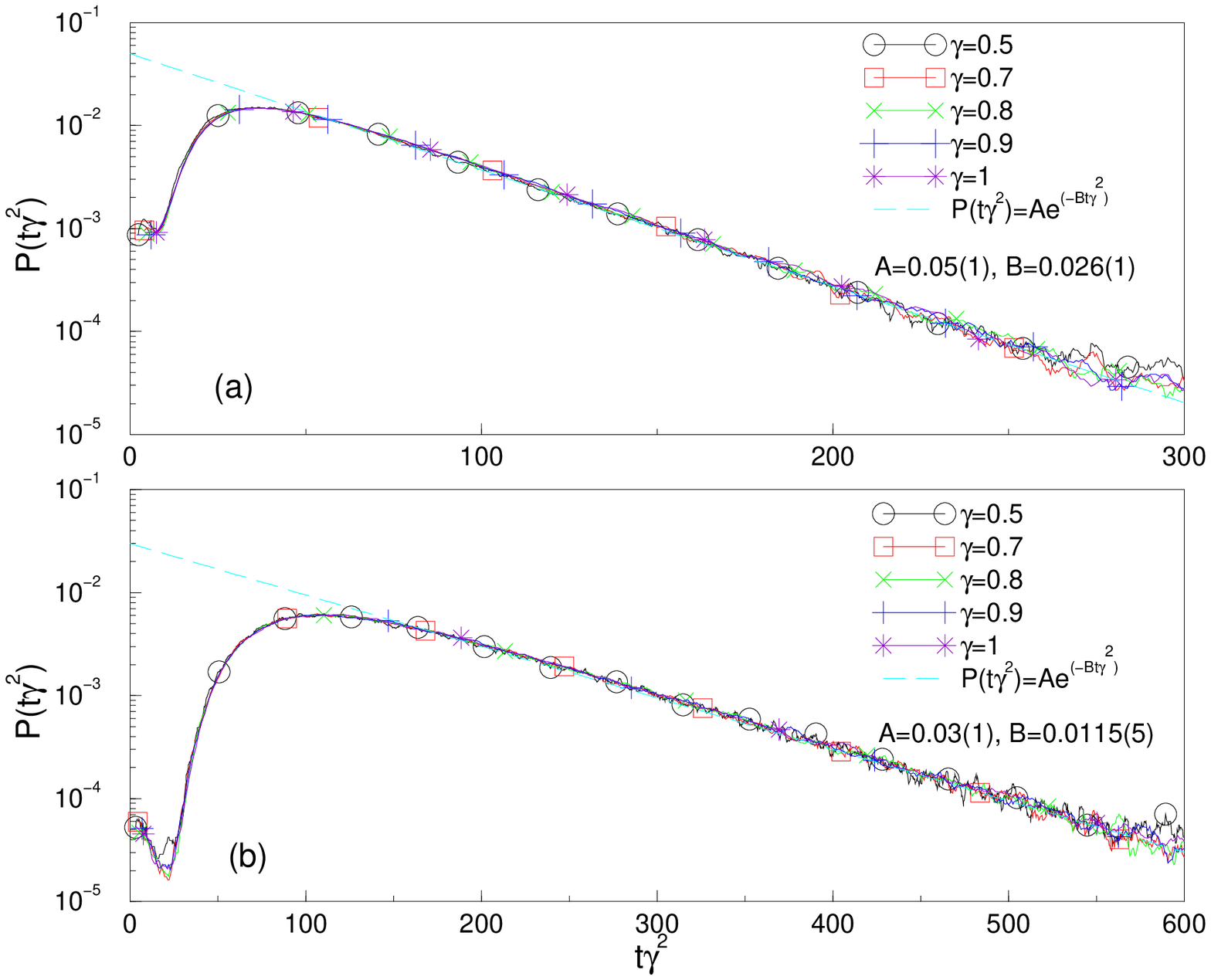,height=12cm,angle=0}}
\caption{Different density functions for the lasting probability 
for different values of $\gamma$. In the picture above a threshold of $5\%$ of 
aggregate wealth
has been adopted to claim traders' extinction. In the picture below we used a threshold 
ten times lower. This change affects only the scale of time and the parameters of the fit, 
but it does not affect the form of the distribution.}
\label{fig1}
\end{figure}

\begin{figure}
\centerline{\psfig{file=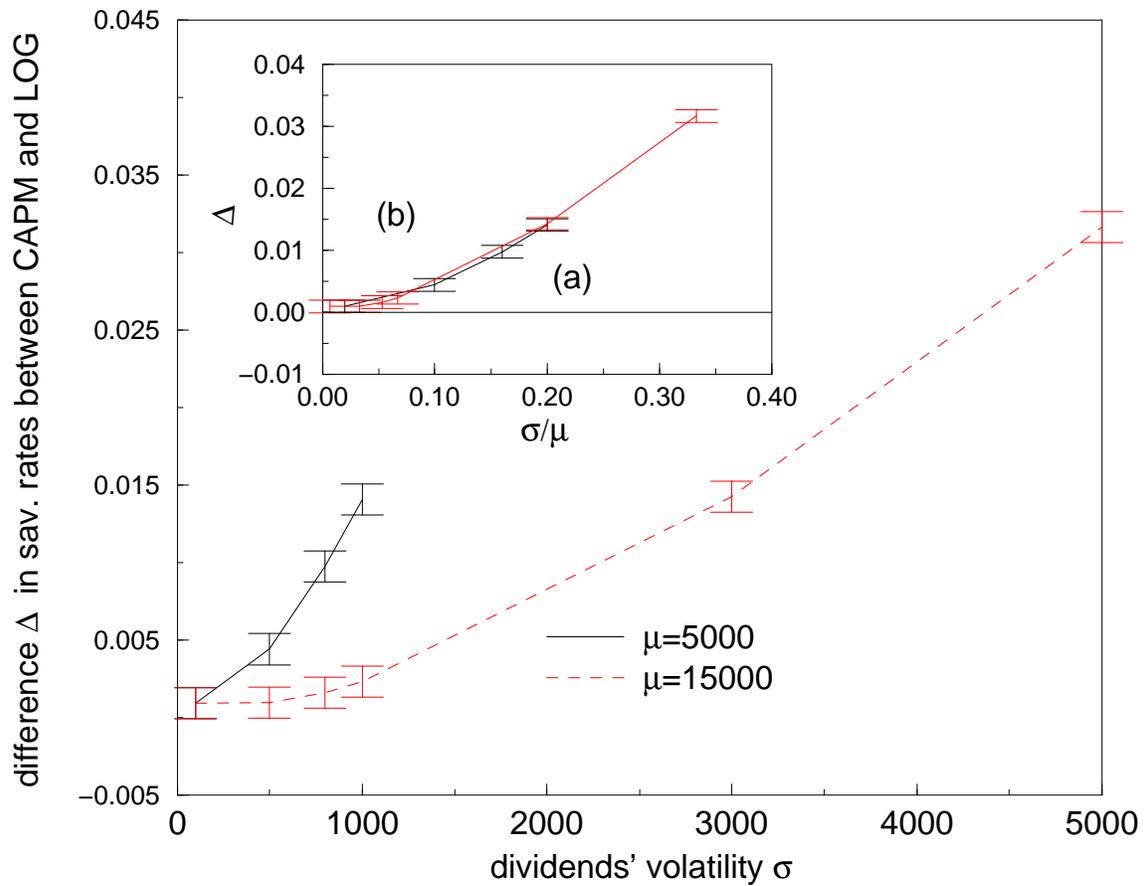,height=12cm,angle=0}}
\caption{Values of $\Delta$ (the maximum difference between the savings rates 
of $CAPM$ and $LOG$ that still allows $LOG$ traders to dominate), with respect to the 
dividends' volatility $\sigma$.}
\label{fig2}
\end{figure}
\end{document}